\documentclass[]{aa}
\usepackage{siunitx}
\usepackage{xcolor}
\usepackage{dsfont}
\usepackage{esint}
\usepackage{physics}
\numberwithin{equation}{section}

\begin{document}

\title{Investigating the damping rate of phase-mixed Alfv\'en waves}
\author{A.~P.~K.~Prokopyszyn \and A.~W.~Hood}
\institute{School of Mathematics and Statistics, University of St Andrews, St Andrews, Fife, KY16 9SS, U.K. 
\\
\email{apkp@st-andrews.ac.uk}
}

\abstract
{This paper investigates the effectiveness of phase mixing as a coronal heating mechanism. A key quantity is the wave damping rate, $\gamma$, defined as the ratio of the heating rate to the wave energy.}
{We investigate whether or not laminar phase-mixed Alfv\'en waves can have a large enough value of $\gamma$ to heat the corona. We also investigate the degree to which the $\gamma$ of standing Alfv\'en waves which have reached steady-state can be approximated with a relatively simple equation. Further foci of this study are the cause of the reduction of  $\gamma$ in response to leakage of waves out of a loop, the quantity of this reduction, and how increasing the number of excited harmonics affects $\gamma$.}
{We calculated an upper bound for $\gamma$ and compared this with the $\gamma$ required to heat the corona. Analytic results were verified numerically.}
{We find that  at observed frequencies $\gamma$ is too small  to heat the corona by approximately three orders of magnitude. Therefore, we believe that laminar phase mixing is not a viable stand-alone heating mechanism for coronal loops. To arrive at this conclusion, several assumptions were made. The assumptions are discussed in Section  \ref{sec:justification}. A key assumption is that we model the waves as strictly laminar. We show that $\gamma$ is largest at resonance. Equation \eqref{eq:dr_approx} provides a good estimate for the damping rate (within approximately 10\% accuracy) for resonant field lines. However, away from resonance, the equation provides a poor estimate, predicting $\gamma$ to be orders of magnitude too large. We find that leakage acts to reduce $\gamma$ but plays a negligible role if $\gamma$ is of the order required to heat the corona. If the wave energy follows a power spectrum with slope -5/3 then $\gamma$ grows logarithmically with the number of excited harmonics. If the number of excited harmonics is increased by much more than 100, then the heating is mainly caused by gradients that are parallel to the field rather than perpendicular to it. Therefore, in this case, the system is not heated mainly by phase mixing.}
{}

\titlerunning{Damping rate of phase-mixed Alfv\'en waves}
\authorrunning{A.~P.~K.~Prokopyszyn \& A.~W.~Hood}

\maketitle

\section{Introduction}

The coronal heating problem relates to the question of why the temperature of the corona is over a hundred times hotter than the photosphere; see for example \citet{Klimchuk2006}, \citet{Parnell2012}, \citet{DeMoortel2015} and \citet{Klimchuk2015} for an overview of the coronal heating problem and the open questions that remain to be addressed. The corona being hotter than the chromosphere, conduction is an energy loss mechanism.  Similarly, the corona is optically thin, and therefore radiation is a loss mechanism. Ohmic dissipation of electric currents and viscous dissipation of plasma motions are thought to play a major role in balancing the conductive and radiative losses in the corona \citep{Klimchuk2015}. It is unclear whether Ohmic or viscous heating dominates the other. Moreover, the precise mechanism(s) by which these occur remains an open question. The proposed mechanisms can be split into two categories: reconnection and wave heating mechanisms. This paper focuses on phase mixing, one of the wave heating mechanisms.

Magnetohydrodynamic (MHD) waves are commonplace in the solar atmosphere and have been observed over the last two decades as a consequence of new, improved imaging and spectroscopic instruments (see, e.g. \citet{Tomczyk2007}, \citet{McIntosh2011} and \citet{DeMoortel2012}). A review of the linear behaviour of MHD waves can be found in, for example, \citet{Goossens2011}. The dissipation of Alfv\'en waves has been the basis of many coronal heating models (see review by \citet{Arregui2015} and references therein). It is believed that the main mechanisms by which  Alfv\'en waves are converted into heat are Ohmic and viscous dissipation. Both of these mechanisms are proportional to the current density and vorticity. There are several proposed mechanisms which may be able to dissipate waves at a high enough rate to heat the corona. The mechanism that this paper focuses on is phase mixing as suggested by \citet{Heyvaerts1983}. Phase mixing is the process where gradients perpendicular to the field build-up due to Alfv\'en waves propagating on field lines with a spatial gradient in Alfv\'en travel time. This process leads to neighbouring waves moving out of phase with each other, hence the name phase mixing. Other notable mechanisms are: resonant absorption \citep{Ionson1982}, reflection-driven Alfv\'en wave turbulence (\citet{Hollweg1986}, \citet{vanBallegooijen2011} and \citet{Shoda2019}), turbulence triggered via the tearing mode or Kelvin-Helmholtz instability (\citet{Heyvaerts1983}, \citet{Browning1984}, \citet{Antolin2016} and \citet{Antolin2018}) and coupling with compressive modes (\citet{Kudoh1999} and \citet{Antolin2010}). 

Phase mixing has been researched quite extensively since \citet{Heyvaerts1983} first proposed it as a coronal heating mechanism. For example, \citet{Browning1984} expanded on the Kelvin-Helmholtz analysis and argue that the phase mixing of standing Alfv\'en waves can trigger turbulence, which can lead to a turbulent cascade and enhanced dissipation of wave energy. \citet{Similon1989} and \citet{Howson2019} study phase mixing in a complex magnetic field. \citet{Parker1991} investigated phase mixing in a coronal hole and argues that it is not valid to assume an ignorable coordinate, and therefore that Alfv\'en waves couple to other modes and are not subject to pure phase mixing. \citet{Hood1997} and \citet{Hood2002} investigate phase mixing in coronal holes, and find a self-similar solution which enables them to analyse a more general class of solutions. These latter authors find that a single pulse decays algebraically rather than exponentially. More recently, phase mixing has been investigated in 3D \citep{Magyar2017} and 3D coronal loops (\citet{Pagano2017} and \citet{Pagano2018}). 

Here, we aim to provide a critical analysis of laminar phase mixing in coronal loops. Through our analysis, we aim to assess whether phase mixing can provide a sufficient damping rate, $\gamma$. The damping rate, $\gamma$, is defined as
\begin{equation}
    \label{eq:damping_rate}
    \begin{aligned}
        \gamma &= \frac{\langle\text{Heating rate}\rangle}{2\times\langle\text{Kinetic wave energy}\rangle}, \\
        &= \frac{\langle\oiint_{\partial V}\vec{E}\times\vec{B}/\mu\cdot\vec{dS}\rangle}{\langle\iiint_V\rho u^2dV\rangle},
    \end{aligned}
\end{equation}
where $\langle\rangle$ denotes the time average of either $\vec{E}\times\vec{B}/\mu$, the Poynting flux, or $\rho u^2/2$, the kinetic energy density. We assumed that kinetic and magnetic energy is transported into the corona mainly via Poynting flux from lower layers in the atmosphere, which is then dissipated as heat \citep{Klimchuk2015}. This definition was chosen in part because it can be calculated relatively easily using observational data. The required heating rate has been estimated by, for example, \citet{Withbroe1977}. The amplitude of transverse oscillations in the corona has been observed by, for example, \citet{McIntosh2011}. This definition approximates the more classical definition of the damping rate. To see this, we consider a damped harmonic oscillator with amplitude, $x(t)$, with its motion described by
\begin{equation}
    \ddot{x}+\gamma\dot{x}+\omega_0^2x=0,
\end{equation}
hence, 
\begin{equation}
    \frac{d}{dt}\left(\frac{1}{2}\dot{x}^2+\frac{1}{2}\omega_0^2x^2\right)=-\gamma\dot{x}^2,
\end{equation}
and so the heating rate is given by $\gamma\dot{x}^2$. Therefore,
\begin{equation}
    \begin{aligned}
        \gamma &= \frac{\langle\text{Heating rate}\rangle}{2\times\langle\text{Kinetic energy}\rangle}, \\
        &= \frac{\gamma\langle\dot{x}^2\rangle}{\langle\dot{x}^2\rangle}.
    \end{aligned}
\end{equation}
\citet{Hollweg1984simple,Hollweg1984resonances} uses a very similar idea to $\gamma$ except he uses a quantity called the quality factor, $Q$, from resonance theory, which is approximately given by $Q = \omega / \gamma$, where $\omega$ is the angular frequency of a wave. We focus on using $\gamma$ because it is easier to apply to a system in which multiple harmonics oscillate. \citet{Arregui2015} also stresses the importance of the damping rate, although he does not give a precise definition. He argues, (through an order magnitude analysis) that phase mixing could take too long to reach the required length scales for the heating to become important. In this paper, we show that even if the waves are allowed to evolve to steady state, the damping rate is still too low.

A damping rate of about $10^{-1}\si{.s^{-1}}$ is required to heat coronal loops, which is based on predictions of the required heating rate and observations of velocity fluctuations in the corona. If we assume that the radiative loss function, $\Lambda_{rad}(T)$, is given by
\begin{equation}
    \Lambda_{rad}(T)=\chi T^{-1/2},
\end{equation}
where $T$ is temperature and $\chi=10^{-32}\si{.K^{1/2}.W.m^3}$
and a uniform coronal heating profile, $H_{c}$, then \citet{Priest2014} shows that to maintain a loop with a  uniform pressure, $p_0$, requires a heating rate of
\begin{equation}
    H_c = \frac{7}{8}\frac{\chi}{k_B^2}\frac{p_0^2}{T_{max}^{5/2}},
\end{equation}
where $k_B$ is the Boltzmann constant and $T_{max}$ is the maximum temperature. Here, $T_{max}=10^6\si{.K}$ and $p_0=10^{-2}\si{.Pa}$ gives $H_c\approx10^{-5}\si{.W.m^{-3}}$, and this approximately agrees with \citet{Withbroe1977} for a 100 \si{Mm} loop. \citet{McIntosh2011} and \citet{McIntosh2012} observe amplitudes in the quiet sun of around $20\si{.km.s^{-1}}$ and in active regions of around $5\si{.km.s^{-1}}$. If we assume an amplitude of $u_0=10\si{.km.s^{-1}}$ and a density of $\rho_0=10^{-12}\si{.km.m^{-3}}$, this gives a kinetic energy density of $\rho_0u_0^2/2=10^{-4}\si{.J.m^{-3}}/2$. Therefore, the required damping rate is approximately $H_c/(\rho_0u_0^2)=10^{-1}\si{.s^{-1}}$. 

This paper does not investigate the origin of the coronal waves. We generate the waves using a driver at the edge of our domain.  The origin of coronal waves remains an open question. The exponential density and the steep jump in density at the transition region make it difficult for Alfv\'en waves generated at the photosphere to enter the corona \citep{Cranmer2005}. \citet{Hollweg1984resonances} suggests resonances in coronal loops and spicules provide enough energy flux to the corona. \citet{Cally2011} and \citep{Hansen2012} suggest that mode conversion from fast waves to Alfv\'en waves at the transition region enables sufficient energy flux to enter the corona. It is also possible that the corona itself generates Alfv\'en waves via reconnection \citep{Cranmer2018}.

This paper is structured as follows: Section \ref{sec:equations} shows the equations which we solve in this paper and then discusses some of the simplifications which were used in deriving these equations; in particular, how the Braginskii viscous stress tensor can be simplified. In Section \ref{sec:closed_loop} and all subsequent sections, we consider a closed-loop. The goal of Section \ref{sec:closed_loop} is to calculate the damping rate of standing Alfv\'en waves which have reached steady state and then to show it can be approximated with a relatively simple equation. Section \ref{sec:leakage} discusses the effects of allowing waves to leak out of the loop. Finally, Section \ref{sec:multiple_harmonics} considers the effects of exciting multiple harmonics. Results from Section \ref{sec:multiple_harmonics} are used in the conclusions to deduce that laminar phase mixing is not a viable stand-alone heating mechanism in coronal loops.

\section{Equations}
\label{sec:equations}

\begin{figure}
    \centering
    \includegraphics[width=0.49\textwidth]{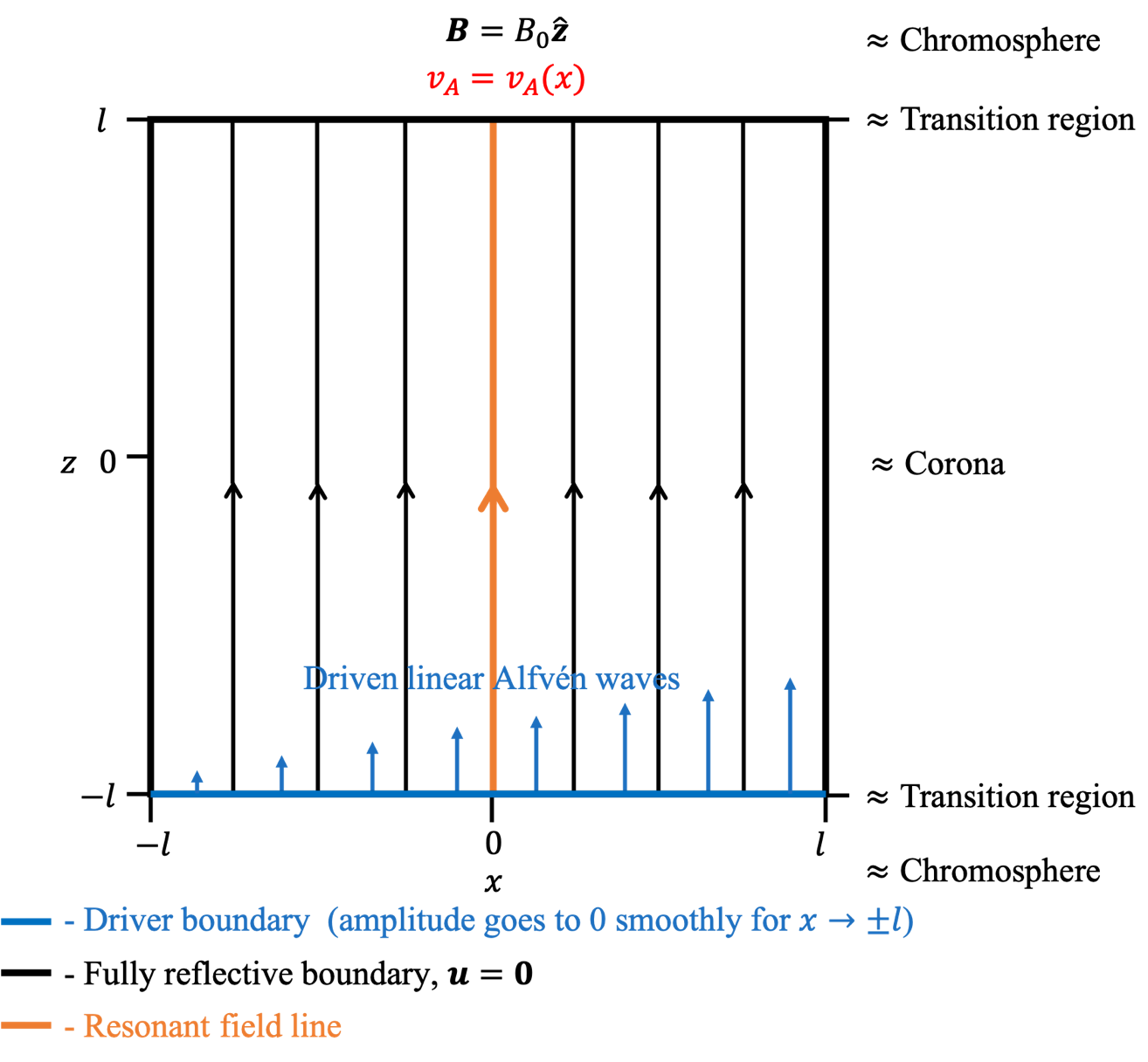}
    \caption{Diagram of our model, showing a set of closed coronal loops (vertical black lines) with  a driver imposed at the bottom boundary (blue). The gradient in height of the blue arrows indicates there is a gradient in Alfv\'en speed.}
    \label{fig:model_diagram}
\end{figure}

Our model is illustrated in Figure \ref{fig:model_diagram}. This section presents the equations that we solve. Section \ref{sec:justification} discusses some of the simplifications which were made to arrive at these equations. The variables we consider are the magnetic field,
\begin{equation}
    \label{eq:magnetic_field}
    \vec{B}=B_0\vec{\hat{z}}+b(x,z,t)\vec{\hat{y}},
\end{equation}
the velocity,
\begin{equation}
    \label{eq:velocity}
    \vec{u}=u(x,z,t)\vec{\hat{y}},
\end{equation}
and the density,
\begin{equation}
    \rho=\rho(x).
\end{equation}
The equations are the linearised momentum equation,
\begin{equation}
    \label{eq:momentum}
    \frac{\partial u}{\partial t}=\frac{B_0}{\mu\rho(x)}\frac{\partial b}{\partial z}+\nu\pdv[2]{u}{x},
\end{equation}
and the linearised induction equation,
\begin{equation}
    \label{eq:induction}
    \frac{\partial b}{\partial t}=B_0\frac{\partial u}{\partial z}+\eta\pdv[2]{b}{x},
\end{equation}
where the permeability of free space is denoted with $\mu$, the viscosity coefficient, $\nu$, is given by
\begin{equation}
\nu=6.63\times10^{-17}\frac{T^{5/2}}{\rho\log \Lambda}(\omega_p\tau_p)^{-2}\si{.m^{2}.s^{-1}}    
,\end{equation}
and the magnetic diffusivity coefficient, $\eta$, is given by
\begin{equation}
    \eta=5.2\times10^{7}\log\Lambda\, T^{-3/2}\si{.m^{2}.s^{-1}},
\end{equation} 
where $\omega_p$ is the proton cyclotron frequency given by
\begin{equation}
    \omega_p=\frac{eB}{m_p},
\end{equation}
with $e$ the elementary charge and $m_p$ the mass of a proton. Here, $\tau_p$ is the proton-proton collision time, given by
\begin{equation}
    \tau_p=2.8\times10^{-20}\frac{T^{3/2}}{\rho\log \Lambda},
\end{equation}
where $\log\Lambda$ is the Coulomb logarithm $\approx22$ in coronal conditions \citep{Priest2014}. Substituting $T=10^6\si{.K}$, $B=10^{-3}\si{.T}$ and $\rho=10^{-12}\si{.m^{-3}}$  gives $\omega_p\tau_p\approx10^5$. We note, $\omega_p \tau_p \gg 1$ corresponds to the strong field limit which means that conduction and viscosity are highly anisotropic. The implications of this are discussed further in Section \ref{sec:justification}. We have neglected spatial derivatives of the viscosity coefficient, $\nu$, and magnetic diffusivity, $\eta$,  in favour of derivatives in the velocity and magnetic field perturbations. Equations \eqref{eq:momentum} and \eqref{eq:induction} can be combined to give
\begin{equation}
    \label{eq:wave_equation}
    \begin{aligned}
        \pdv[2]{u}{t}=v_A^2\pdv[2]{u}{z}+(\eta+\nu)\pdv{}{t}\pdv[2]{u}{x},
    \end{aligned}
\end{equation}
provided the products and squares of the viscosity and magnetic diffusivity coefficients can be neglected, where $v_A=B_0/\sqrt(\mu\rho)$ is the Alfv\'en speed. 

\subsection{Justification}
\label{sec:justification}

\begin{figure}
    \centering
    \includegraphics[width=0.49\textwidth]{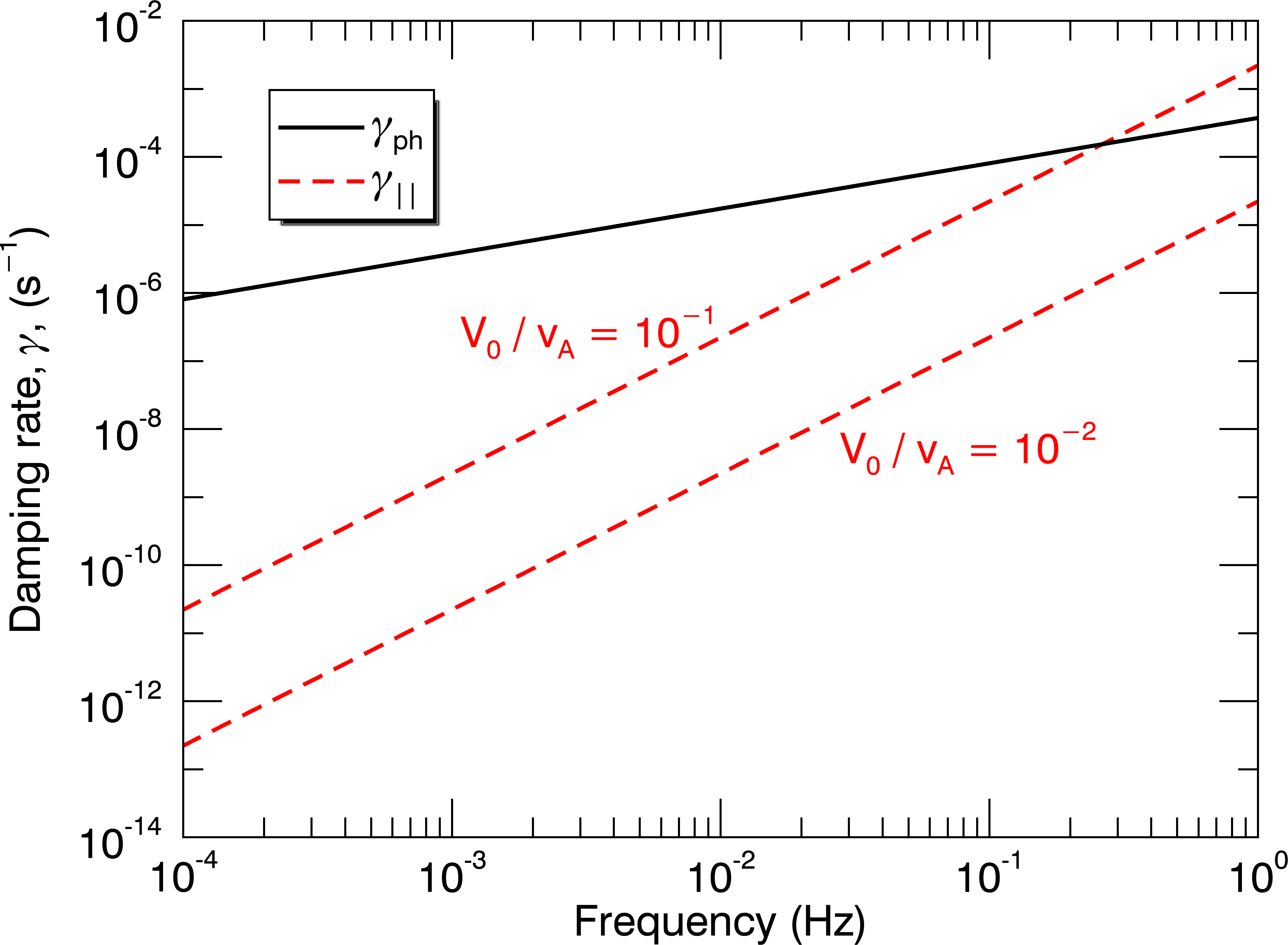}
    \caption{Plots of the estimated damping rates for the case where the only dissipative contribution comes from $\vec{W}^{(0)}$, denoted with $\gamma_{||}$, and for the case where the only dissipative contribution comes from $\vec{W}^{(1)}$, denoted with $\gamma_{ph}$.}
    \label{fig:dr_visc_vs_phase}
\end{figure}

Equations \eqref{eq:magnetic_field} and \eqref{eq:velocity} show that we assume an invariant direction. This approximation may be valid in, for example, coronal arcades or coronal loops which could be approximately invariant along the azimuthal direction. An important consequence of keeping an invariant direction in our system is that it stabilises to the Kelvin-Helmholtz instability and tearing instability. 
To trigger an instability, a disturbance needs to have a wave vector, $\vec{k}$, that satisfies
\[\vec{k}\cdot\vec{B}=k_yb + k_zB_0=0.\]
Our boundary conditions ensure $k_z\ne0$ and since $\pdv*{}{y}=0,$ this ensures that $k_y=0$. Therefore,
\[\vec{k}\cdot\vec{B}\ne 0\]
for all possible disturbances and so instabilities are not triggered in our system.

Equation \eqref{eq:magnetic_field} shows that we model the field lines as straight. In general, the coronal field is curved, especially in coronal arcades, but the concepts outlined here can still be applied. The wave equation describes linear Alfv\'en waves for a potential field in a field-aligned coordinate system. Therefore, there exists a straight field with a density structure which will reproduce the dynamics of an arcade field. 

For analytical progress, we assume the Alfv\'en speed is uniform along the field lines in the corona. In reality, coronal loops are stratified due to gravity and the field strength may vary as well. Using a uniform loop instead of a non-uniform loop has two key effects: The first effect is that it means the wavelength of our waves do not change as they propagate along the loop. This may affect the phase mixing as shorter-wavelength waves form shorter perpendicular length scales more quickly. However, if our uniform Alfv\'en speed represents the average Alfv\'en speed of a non-uniform loop, then the time-averaged heating rates will not differ significantly. The second effect is that a non-uniform Alfv\'en speed can give rise to reflection. If the wavelength of a wave is shorter than the length-scale of the Alfv\'en speed variations, then to a good approximation the reflection is negligible \citep{Leroy1980}. 
In the corona, we estimate the pressure scale height to be approximately $60\si{.Mm}$. Therefore, we argue that for waves with a wavelength smaller than this (say $<50\si{.Mm}$) the reflection within the corona itself can be ignored. We note that a semi-circular loop of length $100\si{.Mm}$ has a vertical height of approximately $32\si{.Mm}$ at the apex. In this paper, we present results with a wavelength varying from $10\si{.Mm}$ to $200\si{.Mm}$. Waves with a longer wavelength than the pressure scale height are presented as we believe their results are still informative.

Our equations to model the viscosity and resistivity are much simpler than that outlined in \citet{Braginskii1965}; the remainder of this paragraph explains how we justify our expressions. As pointed out by \citet{Russell2019} (in review), linearising the viscosity can result in the predicted viscosity being orders of magnitude smaller than it otherwise would be. The viscous force is modelled as the divergence of the Braginskii stress tensor, $\vec{\sigma}_{brag}$, given here in the same form as in \citet{MacTaggart2017},
\begin{equation}
    \begin{aligned}
    \vec{\sigma}_{brag}&=\eta_0\vec{W}^{(0)}+\eta_1\vec{W}^{(1)}+\eta_2\vec{W}^{(2)}\\
    &-(\eta_3\vec{W}^{(3)}+\eta_4\vec{W}^{(4)}),
    \end{aligned}
\end{equation}
where
\begin{equation}
    \vec{W}^{(0)}=\tfrac{3}{2}(\vec{W}\vec{\hat{B}}\cdot\vec{\hat{B}})(\vec{\hat{B}}\otimes\vec{\hat{B}}-\tfrac{1}{3}\vec{I}),
\end{equation}
\begin{equation}
    \begin{aligned}
    \vec{W}^{(1)}&=(\vec{I}-\vec{\hat{B}}\otimes\vec{\hat{B}})\vec{W}(\vec{I}-\vec{\hat{B}}\otimes\vec{\hat{B}}) \\
    &+\tfrac{1}{2}(\vec{W}\vec{\hat{B}}\cdot\vec{\hat{B}})(\vec{I}-\vec{\hat{B}}\otimes\vec{\hat{B}}),
    \end{aligned}
\end{equation}
\begin{equation}
    \begin{aligned}
    \vec{W}^{(2)}&=(\vec{I}-\vec{\hat{B}}\otimes\vec{\hat{B}})\vec{W}(\vec{\hat{B}}\otimes\vec{\hat{B}}) \\
    &+(\vec{\hat{B}}\otimes\vec{\hat{B}})\vec{W}(\vec{I}-\vec{\hat{B}}\otimes\vec{\hat{B}}),
    \end{aligned}
\end{equation}
\begin{equation}
    \vec{W}^{(3)}=\tfrac{1}{2}\vec{Z}\vec{W}(\vec{I}-\vec{\hat{B}}\otimes\vec{\hat{B}})-\tfrac{1}{2}(\vec{I}-\vec{\hat{B}}\otimes\vec{\hat{B}})\vec{W}\vec{Z},
\end{equation}
\begin{equation}
    \vec{W}^{(4)}=(\vec{Z}\vec{W}\vec{\hat{B}})\otimes\vec{\hat{B}}+\vec{\hat{B}}\otimes(\vec{Z}\vec{W}\vec{\hat{B}}),
\end{equation}
with 
\begin{equation}
    \vec{W}=\vec{\nabla}\vec{u}+(\vec{\nabla}\vec{u})^T-\tfrac{2}{3}(\vec{\nabla}\cdot\vec{u})\vec{I},
\end{equation}
where \vec{Z} is the tensor with components $Z_{ij}=\epsilon_{ikj}b_k$, where $\epsilon_{ikj}$ are components of the Levi-Civita symbol.
The viscosity coefficients (see \citet{Braginskii1965}) are given by
\begin{equation}
    \begin{aligned}
    &\eta_0=2.21\times10^{-16}\frac{T^{5/2}}{\log \Lambda}\si{.kg.m^{-1}.s^{-1}}, \\
    &\eta_1=\tfrac{3}{10}\eta_0(\omega_p\tau_p)^{-2}, \quad \eta_2=4\eta_1,\\
    &\eta_3=\tfrac{1}{2}\eta_0(\omega_p\tau_p)^{-1},\quad\eta_4=2\eta_3,
    \end{aligned}
\end{equation}
assuming the strong field limit $(\omega_p\tau_p\gg1)$. We aim to show that $\eta_0\vec{W}^{(0)}$, $\eta_2\vec{W}^{(2)}$, $\eta_3\vec{W}^{(3)}$, $\eta_4\vec{W}^{(4)}$ can be neglected in favour of $\eta_1\vec{W}^{(1)}$. This is surprising because if $\omega_p\tau_p\approx10^5$ then this means that $\eta_1\approx10^{-10}\eta_0$.

Here we assume that $\vec{u}$ and $\vec{b}$ are footpoint-driven waves on closed loops that have reached steady state such that their amplitudes do not change with time and are of the form,
\begin{equation}
    \vec{u}=V(x)\sin(kz)\cos(\omega t+\phi(x))\vec{\hat{y}},
\end{equation}
\begin{equation}
    \vec{B}=B_0\frac{V(x)}{v_A}\cos(kz)\sin(\omega t+\phi(x))\vec{\hat{y}}+B_0\vec{\hat{z}},
\end{equation}
where $k$ denotes the wave number. If we assume the viscosity coefficients are uniform then the viscous heating is given by $\vec{\sigma}_{brag}:\vec{\nabla u}$. The heating contributions from each term, with $\vec{u}$ and $\vec{B}$ given by equations \eqref{eq:magnetic_field} and \eqref{eq:velocity} is given by:
\begin{equation}
    \begin{aligned}
    \eta_0&\vec{W}^{(0)}:\vec{\nabla u}=\eta_0\frac{b^2}{B_0^2}\left(\pdv{u}{z}\right)^2+O\left(\frac{b^4}{B_0^4}\right), \\
    &=\frac{3}{4}\eta_0k^2\frac{V^4}{v_A^2}\cos^4(kz)\sin^2[2(\omega t+\phi)]+O\left(\frac{b^4}{B_0^4}\right)
    \end{aligned}
\end{equation}
\begin{equation}
    \eta_1\vec{W}^{(1)}:\vec{\nabla u}=\eta_1\left(\pdv{u}{x}\right)^2+O\left(\frac{b^2}{B_0^2}\right),
\end{equation}
\begin{equation}
    \eta_2\vec{W}^{(2)}:\vec{\nabla u}=\eta_2\left(\pdv{u}{z}\right)^2+O\left(\frac{b^2}{B_0^2}\right),
\end{equation}
\begin{equation}
    \eta_3\vec{W}^{(3)}:\vec{\nabla u}=\eta_4\vec{W}^{(4)}:\vec{\nabla u}=0.
\end{equation}
We expect  $v/v_A$ to be approximately in the range $10^{-2}$ to $10^{-1}$ (\citet{McIntosh2011} and \citet{McIntosh2012}), and therefore we can neglect the heating from $\vec{W}^{(2)}$ in favour of $\vec{W}^{(0)}$ since we estimate $\eta_0\approx10^{10}\eta_2$.
The goal now is first to calculate the damping rate for our wave in the case where the only dissipative contribution comes from $\vec{W}^{(0)}$. We then compare this with the damping rate for the case where the only dissipative contribution comes from $\vec{W}^{(1)}$ and this can be used to estimate where $\vec{W}^{(1)}$ can be neglected in favour of $\vec{W}^{(0)}$. The average wave energy on a field line is given by
\[\left\langle\int_0^L \rho u^2 dz\right\rangle=\frac{1}{4}\rho L V^2.\]
The average heating rate for the case where the only dissipative contribution comes from $\vec{W}^{(0)}$ along a field line of length $L$ is given by
\[\left\langle \int_0^L  \eta_0\vec{W}^{(0)}:\vec{\nabla u} dz\right\rangle=\frac{9}{64}L\eta_0k^2\frac{V^4}{v_A^2}\]
to leading order. Therefore, the damping rate is given by
\[\begin{aligned}
\gamma_{||}=\frac{9}{16}\frac{\eta_0}{\rho}k^2\frac{V^2}{v_A^2}
\end{aligned}\]
to leading order, denoted with the symbol $\gamma_{||}$ as the heating is only dependent on gradients parallel to the magnetic field.
We note that 
\[\vec{\nabla}\cdot\vec{W}^{(1)}=\eta_1\pdv[2]{u}{x}\vec{\hat{y}}+O\left(\frac{b^2}{B_0^2}\right),\]
and therefore to find the damping rate of our wave in the case where the only dissipative contribution comes from $\vec{W}^{(1)}$ we simply need to use standard phase mixing results \citep{Heyvaerts1983}. In Sect. \ref{sec:closed_loop} we show that the damping rate of a phase mixed Alfv\'en waves can be closely approximated by
\[\gamma_{ph}=\left(\frac{4}{3}\frac{\eta_1}{\rho}\frac{\omega^2}{v_A^2}(\nabla_\perp v_A)^2\right)^{1/3},\]
where $\nabla_{\perp}$ denotes the gradient in Alfv\'en speed perpendicular to the field. Both damping rates are plotted as a function of frequency, $f$, (where $k = 2\pi f / v_A$) in Figure \ref{fig:dr_visc_vs_phase}. To make this plot, the following parameters were used: $\eta_0/\rho=10^{10}\si{.m^2.s^{-1}}$, $\eta_1/\rho=1\si{.m^2.s^{-1}}$, $v_A=1\si{.Mm.s^{-1}}$ and $\nabla_\perp v_A=1\si{.s^{-1}}$ . It can be seen that $\gamma_{ph}\gg\gamma_{||}$ for $f\le10^{-1}\si{.Hz}$ and so in this parameter space we can neglect $\vec{W}^{(0)}$ in favour of $\vec{W}^{(1)}$. It is worth noting that in equation \eqref{eq:momentum} we have neglected the possibility of mode coupling from Alfv\'en waves to magnetoacoustic waves and pondermotive wings \citep{Verwichte1999} and this could enhance the importance of $\vec{W}^{(0)}$. If 
\[\vec{u}=u_x(x,z,t)\vec{\hat{x}}+u_z(x,z,t)\vec{\hat{z}},\]
and 
\[\vec{B}=b_x(x,z,t)\vec{\hat{x}}+B_0\vec{\hat{z}},\]
then
\[\vec{W}^{(0)}:\vec{\nabla u}=\frac{1}{3}\left(\pdv{u_x}{x}-2\pdv{u_z}{z}\right)^2+O\left(\frac{b_x}{B_0}\right),\]
which shows that, in general, the $\vec{W}^{(0)}$ term decays magnetoacoustic waves more efficiently than it does Alfv\'en waves as it is proportional to $(b/B_0)^0$ rather than $(b/B_0)^2$. However, considering the non-linear mode coupling of Alfv\'en waves to other wave modes is a sufficiently complex problem that we believe it is best left as a stand-alone paper. To simplify the notation and keep it in accordance with that of \citet{Priest2014} we take the viscous force to be given by $\eta_1\vec{\nabla}\cdot\vec{W}^{(1)}$, consider only the leading order term, and set 
\[\nu=\eta_1/\rho.\]

In the induction equation \eqref{eq:induction} we neglect the diffusion term involving derivatives parallel to the background field. We justify this because we expect the perpendicular gradients to be much stronger than the parallel gradients after the wave has phase mixed. According to \citet{Braginskii1965} the parallel and perpendicular conductivities differ only by approximately a factor of two. We expect the gradients perpendicular to the field to be many times greater than parallel gradients as the waves phase mix. Therefore, we only  consider the perpendicular gradients and set
\[\eta=\frac{1}{\mu\sigma_{||}},\]
where $\sigma_{||}$ is the conductivity parallel to the field.

Important conditions \citep{Braginskii1965} for the validity of the transport coefficients are
\begin{equation}
    \label{eq:braginskii_conditions}
    L_\perp\gg r_{T,e},\ L_{||}\gg l_{T,p},
\end{equation}
where $L_\perp$ and $L_{||}$ are the characteristic distances in the directions perpendicular and parallel to the magnetic field, $r_{T,e}$ is the thermal electron gyro radius and $l_{T,p}$, is the proton mean free path. Assuming full ionisation and a purely hydrogen corona, the electron gyro radius is given by
\[r_{T,e}\approx2.2\times10^{-2}\left(\frac{T}{10^6\si{.K}}\right)^{1/2}\left(\frac{10^{-3}\si{.T}}{B}\right)\si{.m}.\]
The proton mean free path can be written as
\[l_{T,p}\approx1.2\times10^5\left(\frac{T}{10^6\si{.K}}\right)^2\left(\frac{10^{-12}\si{.kg.m^{-3}}}{\rho}\right)\si{.m}.\]
\citet{Priest2014} shows that the length scale of phase-mixed Alfv\'en waves perpendicular to the field lines, $2\pi/k_x^*$, can be approximated by
\begin{equation}
    \begin{aligned}
    \frac{2\pi}{k_x^*}&=2\pi\left(\frac{(\eta+\nu)\lambda}{12\pi \nabla_\perp v_A}\right)^{1/3}, \\
    &\approx1.9\left(\frac{\eta+\nu}{1\si{.m^2.s^{-1}}}\frac{1\si{.s^{-1}}}{\nabla_\perp v_A}\lambda\right)^{1/3}.
    \end{aligned}
\end{equation}
Therefore, for our parameter space, the conditions in Equation \eqref{eq:braginskii_conditions} are usually satisfied if the wavelength, $\lambda$, is greater than $100\si{.km}$ with the first condition being easier to satisfy than the second.

\section{Closed loop}
\label{sec:closed_loop}

\citet{Heyvaerts1983} and \citet{Priest2014} find the solution for an Alfv\'en wave on an open field line by assuming a solution of the form,
\begin{equation}
    \label{eq:phase_mixing_soln1}
    u=V(x,z)e^{i(k_zz-\omega t)},
\end{equation}
and assume $V$ is weakly varying in $z$ which means $\partial V / \partial z \ll k_zV$. They find the solution to be
\begin{equation}
    \label{eq:phase_mixing_soln2}
    V=V_0\exp\left[-\frac{(\eta+\nu)k_z^2}{6v_A^3}(\nabla_\perp v_A)^2z^3\right],
\end{equation}
where $\nabla_\perp v_A=dv_A/dx$ denotes the gradient in Aflv\'en speed perpendicular to the field. This solution has a characteristic damping length, $l_{ph}$, given by
\begin{equation}
    l_{ph}=\left(\frac{(\eta+\nu)\omega^2}{6v_A^5}(\nabla_\perp v_A)^2\right)^{-1/3}.
\end{equation}
This gives a timescale, $\tau_{ph}=l_{ph} v_A^{-1}$, given by 
\begin{equation}
    \tau_{ph}=\left(\frac{(\eta+\nu)\omega^2}{6v_A^2}(\nabla_\perp v_A)^2\right)^{-1/3}.
\end{equation}
The goal of this section is to assess whether the phase-mixed damping rate, $\gamma_{ph}$, as defined in equation \eqref{eq:damping_rate} can be approximated with $2\tau_{ph}^{-1}$, given by
\begin{equation}
    \begin{aligned}
        \label{eq:dr_approx}
        \gamma_{ph} =\left(\frac{4}{3}\frac{(\eta+\nu)\omega^2}{v_A^2}(\nabla_\perp v_A)^2\right)^{1/3}.
    \end{aligned}
\end{equation}
To do this, we extend the open  field solution to
a closed-loop of  \citet{Heyvaerts1983}, as in Figure \ref{fig:model_diagram}. We note that \citet{Heyvaerts1983} have already derived a solution for a closed loop, but we believe the form of  our solution to be more useful for our problem. We derive our formula via a different approach to that of \citet{Heyvaerts1983}; they make use of Green's function whereas we use a method of images approach. In Section \ref{sec:analytic_soln} we calculate a solution analytically for the damping rate at steady state. We then verify our calculation of the Poynting flux numerically in Section \ref{sec:numerical_soln}. Finally, in Section \ref{sec:damping_rate} we discuss how well equation \eqref{eq:dr_approx} approximates the damping rate.

\subsection{Analytic solution}
\label{sec:analytic_soln}

The goal here is to extend Equations \eqref{eq:phase_mixing_soln1} and \eqref{eq:phase_mixing_soln2}  for a confined domain, as in Figure \ref{fig:model_diagram} (but with the complex exponential replaced with a sine function). We consider a domain in which a factor, $R$, of the wave amplitude reflects at the transition region boundary, where the boundaries are given by
\[-l\le z\le l.\]
We note that $R<1$ denotes the amplitude reflection coefficient which is not (in general) the same as the energy reflection coefficient, $R_E$. We solve the problem using a  method of images approach, and the full solution for $u$ is
\begin{equation}
    \begin{aligned}
        \label{eq:full_soln}
        u&=u_0\sum_{k=0}^\infty(-1)^kR^k\exp[-(z_k/l_{ph})^3]H(\theta_k)\sin(\omega\theta_k), \\
        &=u_0\sum_{k=0}^m(-1)^kR^k\exp[-(z_k/l_{ph})^3]H(\theta_k)\sin(\omega\theta_k),
    \end{aligned}
\end{equation}
where $H()$ denotes the Heaviside step function,
\begin{equation}
    \theta_k=t-(-1)^k\frac{z}{v_A}-\frac{2k+1}{v_A}l,
\end{equation}
\begin{equation}
    \begin{aligned}
        z_k=(-1)^kz+(2k+1)l
    \end{aligned}
,\end{equation}
\begin{equation}
    m=\left\lfloor\frac{tv_A}{L}\right\rfloor.
\end{equation}
Here, $\lfloor\rfloor$ denote the floor function, namely the integer part to the real number.
Using the trig identity,
\begin{equation}
    \sin(\omega \theta_k)=\sin(\omega t)\cos(\omega z_k / v_A)-\cos(\omega t)\sin(\omega z_k / v_A),
\end{equation}
the steady-state solution, that is the solution for $t\rightarrow\infty$ to equation \eqref{eq:full_soln}, is given by
\begin{equation}
    \label{eq:u_steady_state}
    u=u_0[A\sin(\omega t)+B\cos(\omega t)],
\end{equation}
where $A$ is given by
\begin{equation}
    \label{eq:A}
    A=\sum_{k=0}^\infty(-1)^kR^k\exp[-(z_k/l_{ph})^3]\cos(\omega z_k / v_A),
\end{equation}
and $B$ is given by
\begin{equation}
    \label{eq:B}
    B=-\sum_{k=0}^\infty(-1)^kR^k\exp[-(z_k/l_{ph})^3]\sin(\omega z_k / v_A).
\end{equation}

The solution for the magnetic field perturbation, $b$, is very similar to that of the velocity and is,
\begin{equation}
    \begin{aligned}
        \label{eq:full_soln_b}
        b&=-u_0\sqrt{\mu\rho}\sum_{k=0}^\infty R^k\exp[-(z_k/l_{ph})^3]H(\theta_k)\sin(\omega\theta_k), \\
        &=-u_0\sqrt{\mu\rho}\sum_{k=0}^mR^k\exp[-(z_k/l_{ph})^3]H(\theta_k)\sin(\omega\theta_k).
    \end{aligned}
\end{equation}
At steady state, this simplifies to
\begin{equation}
    \label{eq:b_steady_state}
    b=-u_0\sqrt{\mu\rho}[C\sin(\omega t)+D\cos(\omega t)],
\end{equation}
where $C$ is given by
\begin{equation}
    \label{eq:C}
    C=\sum_{k=0}^\infty R^k\exp[-(z_k/l_{ph})^3]\cos(\omega z_k / v_A),
\end{equation}
and $D$ is given by
\begin{equation}
    \label{eq:D}
    D=-\sum_{k=0}^\infty R^k\exp[-(z_k/l_{ph})^3]\sin(\omega z_k / v_A).
\end{equation}

The Poynting flux on the boundary can be simplified using Ohm's law and neglecting the small resistive terms gives
\begin{equation}
    \begin{aligned}
        \frac{\vec{E}\times\vec{B}}{\mu}\cdot\vec{\hat{B_0}}&=\eta\vec{j}\times\vec{B}-\frac{1}{\mu}\vec{u}\times\vec{B}\times\vec{B}\cdot\vec{\hat{B_0}}, \\
        &=-B_0ub/\mu.
    \end{aligned}
\end{equation}
Using equations \eqref{eq:u_steady_state} and \eqref{eq:b_steady_state} the average steady-state Poynting flux, $-B_0\langle ub\rangle/\mu$, is given by,
\begin{equation}
    \begin{aligned}
    -B_0\langle ub\rangle/\mu&=\frac{B_0}{2}\sqrt{\frac{\rho}{\mu}}u_0^2(AC+BD), \\
    &=\frac{1}{2}\rho u_0^2 v_A(AC+BD).
    \end{aligned}
\end{equation}
The average steady-state kinetic wave energy density, $\rho\langle u^2\rangle/2$, can be expressed as
\begin{equation}
    \rho\langle u^2\rangle/2=\frac{1}{4}\rho u_0^2(A^2+B^2).
\end{equation}
Therefore, the average steady-state kinetic wave energy, $E_k$, for a field line is
\begin{equation}
    E_k(x)=\int_{-l}^l\frac{1}{4}\rho u_0^2(A^2+B^2)dz.
\end{equation}
Therefore, the damping rate, $\gamma$, for a field line is
\begin{equation}
    \begin{aligned}
        \label{eq:dr}
        \gamma(x) &= \frac{[-B_0\langle u b\rangle/\mu]_{-l}^l}{2E_k} \\
        &=v_A\frac{[AC+BD]_{-l}^l}{\int_{-l}^l(A^2+B^2)dz}.
    \end{aligned}
\end{equation}
We note that for $R=1$ this simplifies to
\begin{equation}
    \label{eq:dr_r=1}
    \gamma(x) = v_A\frac{AC|_{z=-l}}{\int_{-l}^l(A^2+B^2)dz}.
\end{equation}

\subsection{Numerical solution}
\label{sec:numerical_soln}

\begin{figure}
    \centering
    \includegraphics[width=0.3\textwidth]{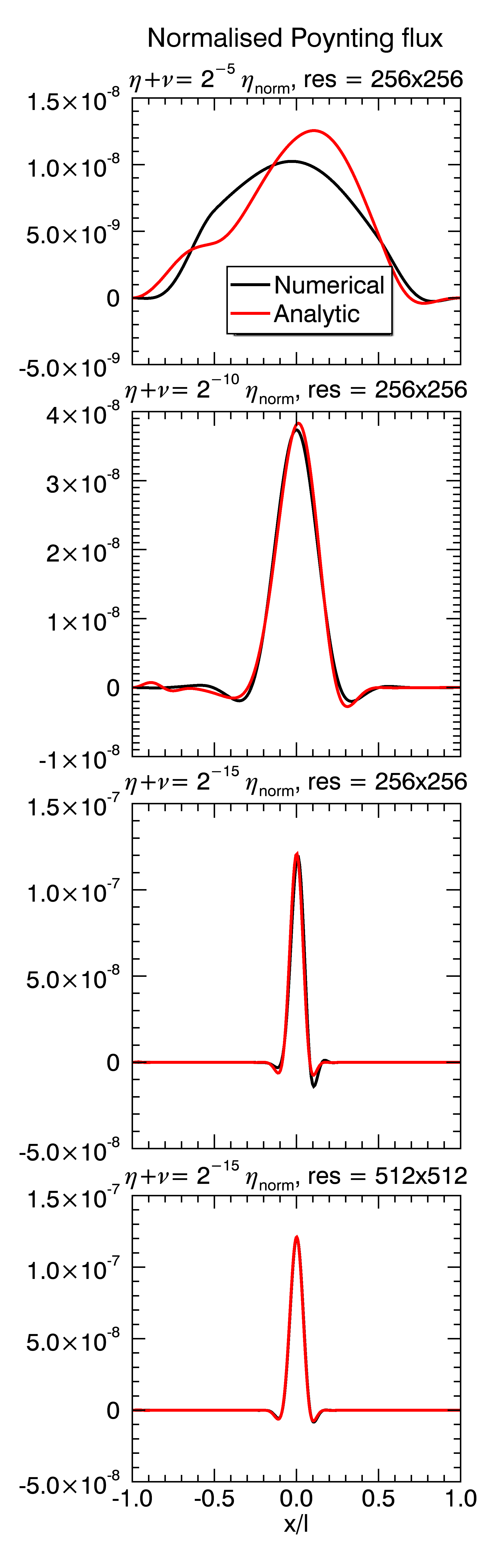}
    \caption{Plots of the numerical and analytic steady-state average Poynting flux for each field line (given by $x=$ a constant) for different values of $\eta+\nu$. Res refers to the resolution used, i.e. the grid size.}
    \label{fig:num_vs_ana_poy_flux}
\end{figure}

The purpose of this section is to demonstrate that Equations \eqref{eq:u_steady_state} and \eqref{eq:b_steady_state} do indeed give accurate solutions provided the damping is weak enough.

The numerical solution is found using the Lare2D code \citep{Arber2001}. Equation \eqref{eq:u_steady_state} and \eqref{eq:b_steady_state} are checked for $R=1$, as this is the easiest boundary condition to impose numerically. The numerical domain is square and is given by,
\begin{equation}
    -l\le x,z\le l.
\end{equation}
In the numerical experiments the Alfv\'en speed, $v_A$, is given by
\begin{equation}
    v_A(x)=v_{A0}(1+x/L),
\end{equation}
where $L=2l$ and $v_{A0}=v_A(0)$. This was chosen as it is the simplest Alfv\'en speed profile with a non-zero gradient. The background magnetic field, $B_0$, is uniform in the $z$ direction. The plasma is initially static. A driver is imposed at the $z=-l$ boundary and has the form
\begin{equation}
    u\vec{\hat{y}}=
    \begin{cases}
    u_0\sin(\omega t)\vec{\hat{y}}, & |x|\le l/2, \\
    u_0\sin(\omega t)\sin^2(\pi x / l)\vec{\hat{y}}, & |x|> l/2,
    \end{cases}
\end{equation}
where $\omega$ is given by
\begin{equation}
    \omega = \pi\frac{v_{A0}}{L}.
\end{equation}
This driver excites resonance at only one location: the orange field line in Figure \ref{fig:model_diagram}, where the fundamental harmonic is excited. In Section \ref{sec:multiple_harmonics} we investigate the effect of using a driver which excites multiple harmonics. Reflective or solid boundary conditions \citep{laney1998} are otherwise imposed on all the boundaries. In other words, $\vec{u}=0$ and $\vec{\hat{n}}\cdot\vec{\nabla}=0$ for all other variables, where $\vec{\hat{n}}$ is a vector normal to the boundary. The boundaries at $z=\pm l$ are designed to simulate the interaction of waves with the transition region. However, perfect reflection is only an approximation; in Section \ref{sec:leakage}, we investigate the effect of using a partially reflective boundary. We note that the code uses isotropic incompressible viscous heating, $\rho\nu\vec{\nabla}u:\vec{\nabla}u$ and isotropic Ohmic heating, $\mu\eta j^2$, where $\vec{j}$ is current density.

Plots of the average steady-state Poynting flux are given in Figure \ref{fig:num_vs_ana_poy_flux}. The Poynting flux peaks at the middle as it is the  fundamental
harmonic of the middle field line, which is excited. As expected, the solutions show better agreement for smaller values of $\eta+\nu$ (provided the resolution is high enough) as this means the \textit{weak damping} assumption from \citet{Heyvaerts1983} is then valid.

\subsection{Damping rate}
\label{sec:damping_rate}

\begin{figure}
    \centering
    \includegraphics[width=0.4\textwidth]{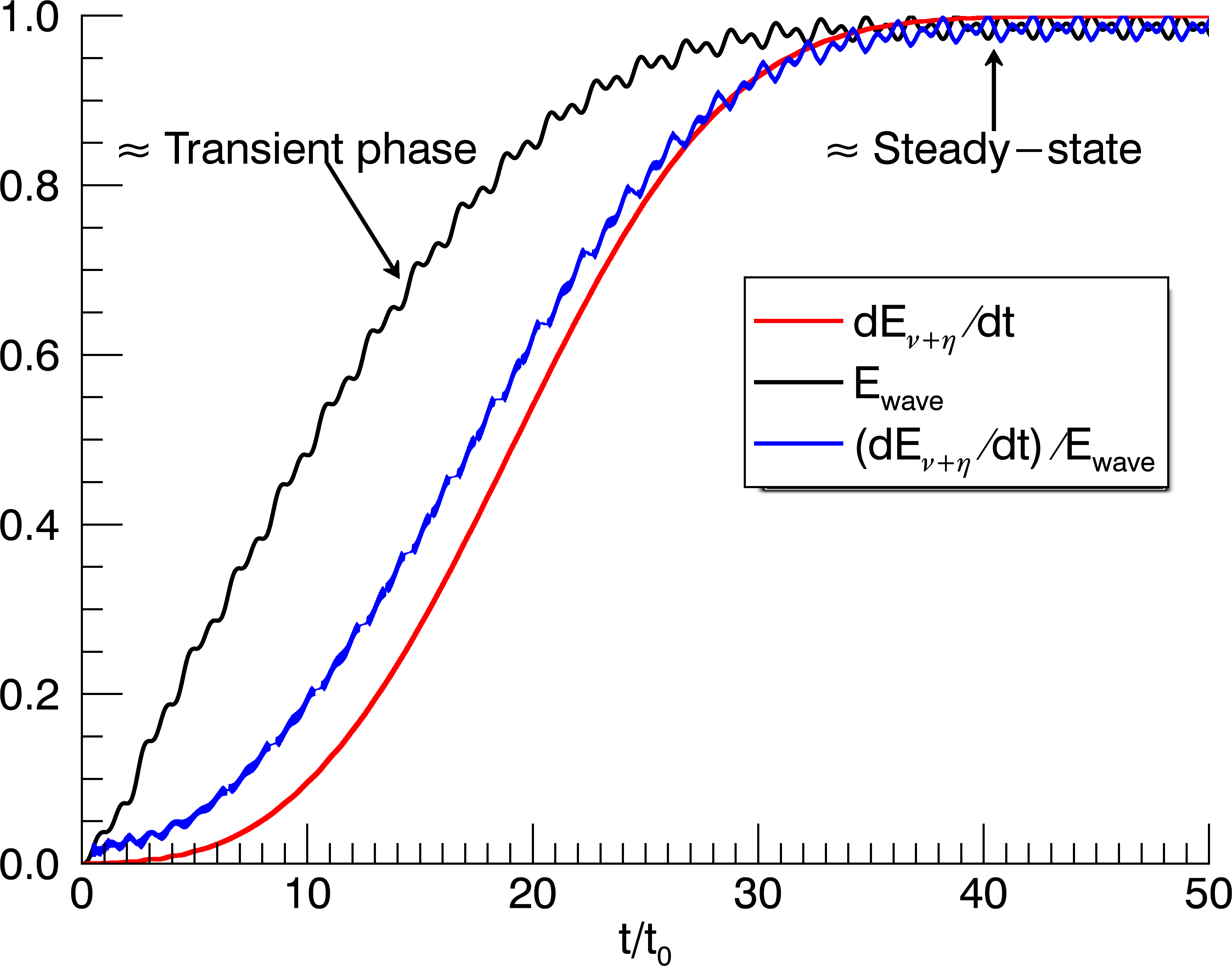}
    \caption{Plots from a numerical experiment showing the rate of change of total Ohmic and viscous energy, $dE_{\eta+\nu}/dt$, total wave energy, $E_{wave}$, and their ratio. Each plot has been normalised by its respective steady-state value and $t_0=L/v_{A0}$.}
    \label{fig:en_int_power_over_en_wave}
\end{figure}

\begin{figure}
    \centering
    \includegraphics[width=0.49\textwidth]{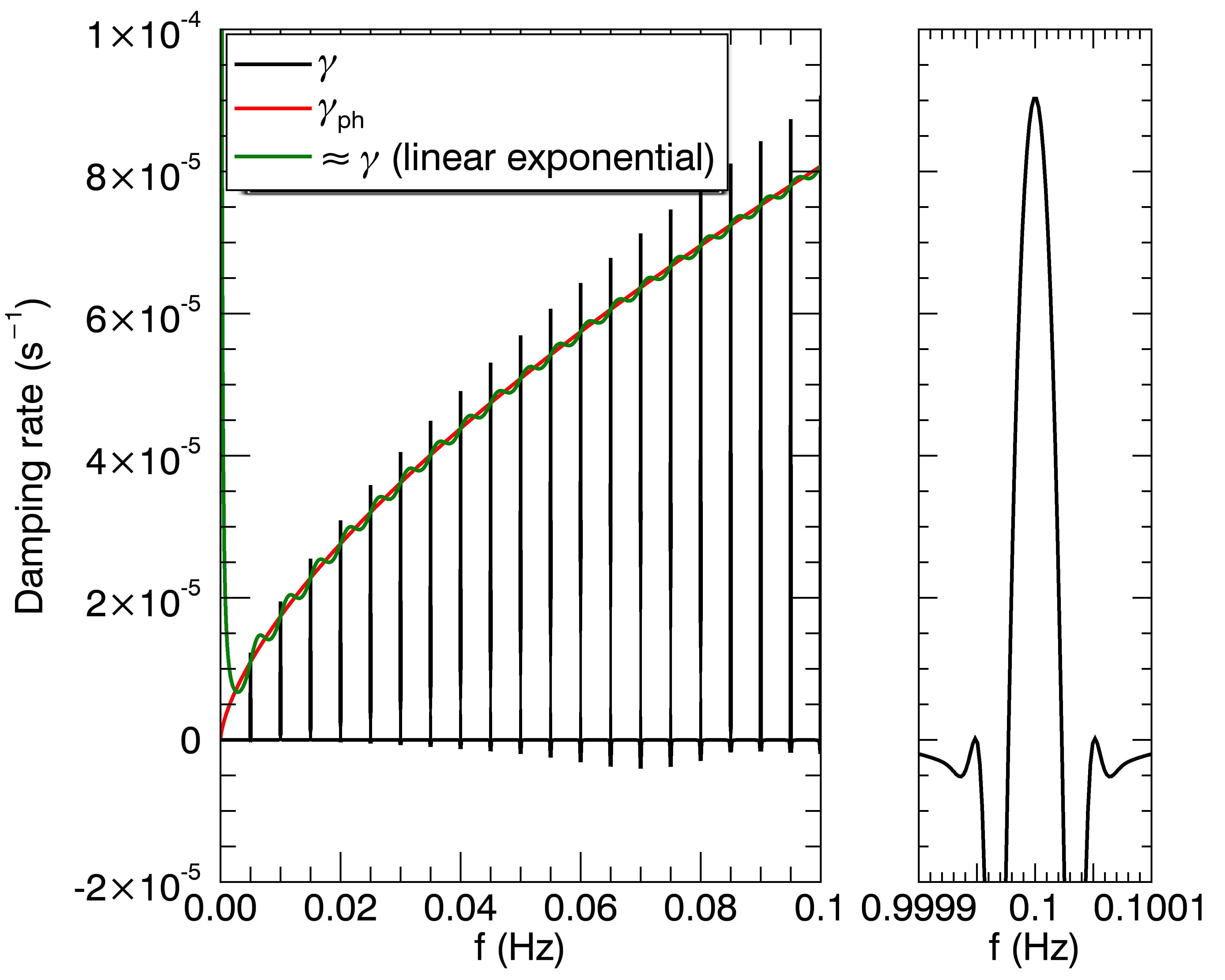}
    \caption{Plot of the steady-state damping rate, $\gamma$, for a field line as a function of the driver frequency. The red curve shows an approximation for the damping rate. The right-hand figure magnifies the black curve in a smaller frequency range to show that the curve is indeed continuous but very steep.}
    \label{fig:dr_vs_freq}
\end{figure}

Figure \ref{fig:en_int_power_over_en_wave} shows plots of the spatially integrated Ohmic and viscous heating, $dE_{\eta+\nu}/dt$, given by
\begin{equation}
    \dv{E_{\eta+\nu}}{t}=\int_{z=-l}^l\int_{x=-l}^l \eta\mu j^2 + \rho \nu \vec{\nabla}\vec{u} : \vec{\nabla}\vec{u} dx dz.
\end{equation}
It also shows the wave energy, $E_{wave}$, where
\begin{equation}
    E_{wave}=\int_{z=-l}^l\int_{x=-l}^l\frac{1}{2}\rho u^2 + \frac{b^2}{2\mu} dxdz.
\end{equation}
Finally, Figure \ref{fig:en_int_power_over_en_wave} shows the ratio, $(dE_{\eta+\nu}/dt)/E_{wave}$; this ratio gives the damping rate, $\gamma$, for our system as a function of time. The figure was produced using data from the numerical experiment described in Section \ref{sec:numerical_soln}, with $\eta+\nu=2^{-15}\eta_{norm}$ and a resolution of $512\times512$, where $\eta_{norm}=2lv_{A0}$. The key result is that the damping rate increases with time and then converges towards a maximum value at steady state. This means that the damping rate at steady state represents an upper bound.

Figure \ref{fig:dr_vs_freq} shows the steady-state damping rate for a field line as a function of frequency. The black curve in Figure \ref{fig:dr_vs_freq} was made using equation \eqref{eq:dr_r=1} with the following parameters: $\eta+\nu=1\si{.m^2.s^{-1}}$,  $v_A=1\si{.Mm.s^{-1}}$, $\nabla_\perp v_A=1\si{.s^{-1}}$, and $L=100\si{.Mm}$ and gives a fundamental frequency of $f_1=5\times10^{-3}\si{.s^{-1}}$. A key result is that the damping rate is significantly larger at resonance and then sharply approaches zero away from resonance (even becoming negative), and therefore the resonant damping rate represents an upper bound. Figure \ref{fig:dr_vs_freq} also shows (in red) our approximation for the damping rate, namely equation \eqref{eq:dr_approx}, which is a good approximation for the natural frequencies. However, there is a noticeable error of less than approximately 10\%. The green curve in Figure \ref{fig:dr_vs_freq} was produced using equation \eqref{eq:dr_r=1}, except the cubic exponential in equations \eqref{eq:A}, \eqref{eq:B}, \eqref{eq:C}, and \eqref{eq:D} were replaced with a linear exponential, namely
\[\exp[-(z_k/l_{ph})^3]\rightarrow\exp[-z_k/l_{ph}].\]
The green curve shows that equation \eqref{eq:dr_approx} is better at predicting the damping rate when there is a linear exponential compared with a cubic exponential. Therefore, the cubic exponential nature of phase mixing reduces the accuracy of equation \eqref{eq:dr_approx} in predicting the damping rate. It is also interesting to note the fact that the damping rate approaches zero (and becomes negative) away from resonance, and this is a result of the cubic nature of phase mixing. If a linear exponential is used, the curve does not approach zero (see green curve) away from resonance.

\section{Leaky loop}
\label{sec:leakage}

\begin{figure}
    \centering
    \includegraphics[width=0.4\textwidth]{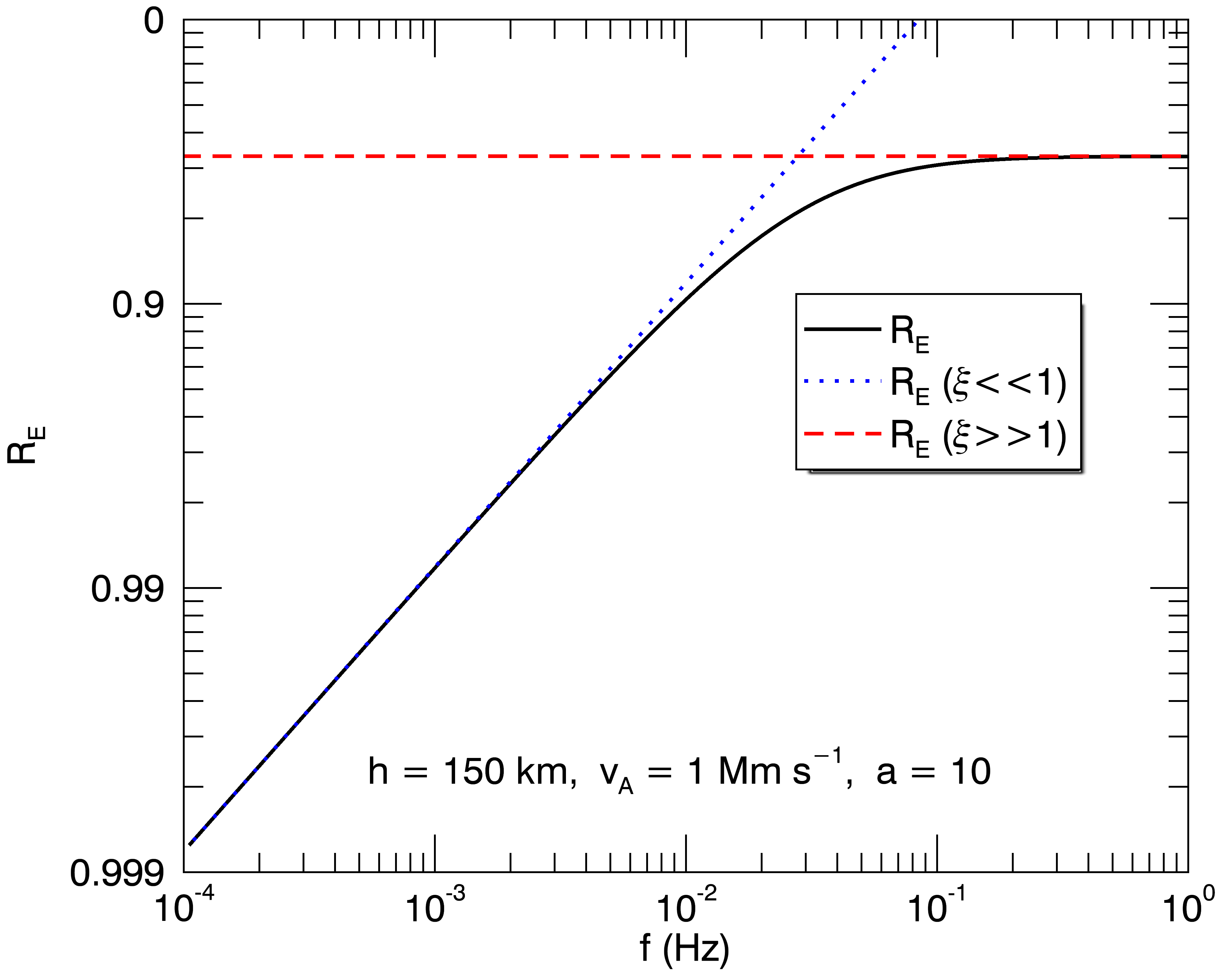}
    \caption{Plot of the energy reflection coefficient, $R_E$, as function of frequency, $f$.}
    \label{fig:reflection_coefficent}
\end{figure}

\begin{figure}
    \centering
    \includegraphics[width=0.4\textwidth]{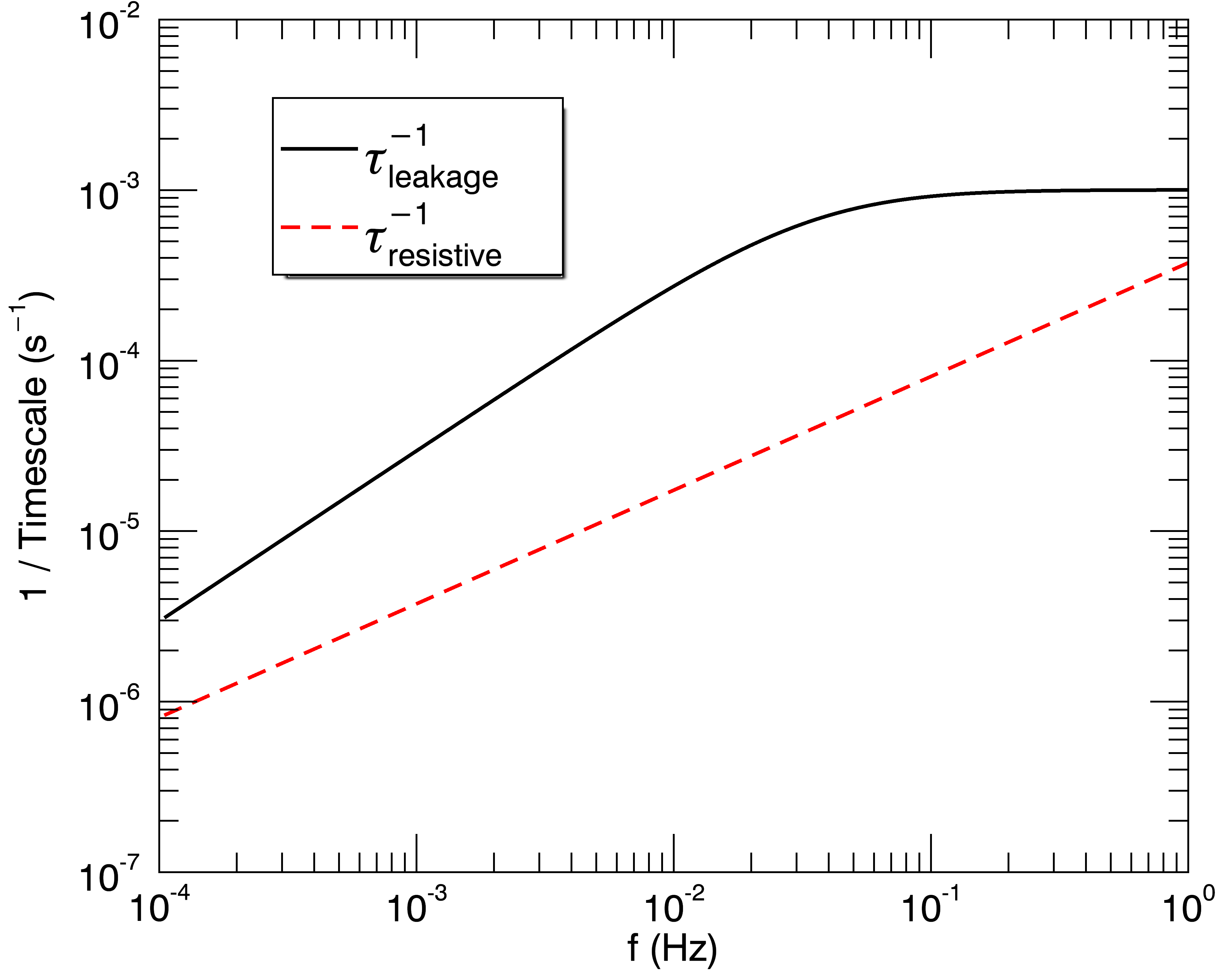}
    \caption{Plots of the leakage timescale, $\tau_{leakagae}$, (see equation \eqref{eq:tau_leakage}) and the resistive timescale, $\tau_{resistive}$, (see equation \eqref{eq:tau_resistive}).}
    \label{fig:leak_vs_resis_timescale}
\end{figure}

\begin{figure}
    \centering
    \includegraphics[width = 0.4\textwidth]{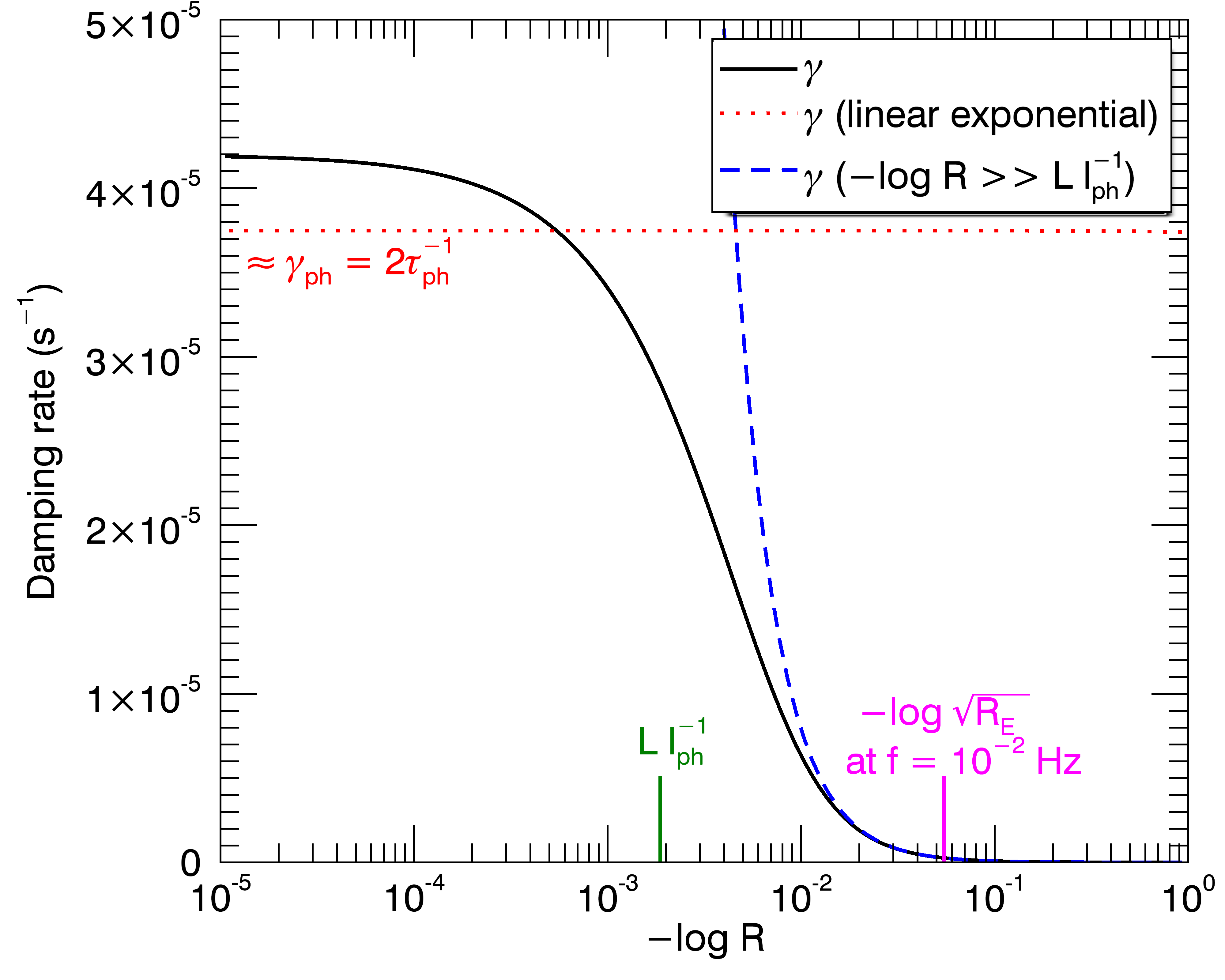}
    \caption{Plots of the damping rate, $\gamma$, as a function of the leakage, $-\log R$.}
    \label{fig:dr_vs_leak}
\end{figure}

Here, we aim to explain why leakage decreases the damping rate and by how much. To do this, we first calculate an expression for the reflection coefficient of Alfv\'en waves at the transition region. After that, we analyse how leakage affects the damping rate.

\citet{Hollweg1984resonances} derives an expression which gives an approximation for the energy reflection coefficient, $R_E$, for Alfv\'en waves at the transition region. To derive this, this latter author modelled the corona as having a uniform Alfv\'en speed and assumed the Alfv\'en speed in the chromosphere varies exponentially in the corona with pressure scale height, $h$. The formula he obtains is
\begin{equation}
    \label{eq:r_e_hollweg}
    R_E=\left|\frac{H_0^{(2)}(\xi)+iH_1^{(2)}(\xi)}{H_0^{(2)}(\xi)-iH_1^{(2)}(\xi)}\right|^2,
\end{equation}
where, 
\begin{equation}
    \label{eq:xi_hollweg}
    \xi=2\frac{h\omega}{v_A}
,\end{equation}
where $v_A$ is the Alfv\'en speed in the corona, $H_0^{(2)}$ denotes the Hankel function of the second kind of  zero order and $H_1^{(2)}$ is the Hankel function of the second kind of first order. We note that the energy transmission coefficient is given by $T_E=1-R_E$. The amplitude reflection coefficient is given by $R = \sqrt{R_E}$. We modify the  \citet{Hollweg1984resonances} model by including a discontinuous jump in Alfv\'en speed from the chromosphere to the corona, which is designed to model the transition region. The behaviour of the wave interaction with the transition region approximates that of a discontinuity provided the wavelength of the wave is greater than the width of the transition region. The Alfv\'en speed increases by a factor, $a$, at the discontinuity and we find that the inclusion of this discontinuity causes the energy reflection coefficient to be instead given by
\begin{equation}
    \label{eq:r_e}
    R_E = \left|\frac{H_0^{(2)}(\xi)+iaH_1^{(2)}(\xi)}{H_0^{(2)}(\xi)-iaH_1^{(2)}(\xi)}\right|^2,
\end{equation}
where,
\begin{equation}
    \label{eq:xi}
    \xi=2a\frac{h\omega}{v_A},
\end{equation}
and this is derived in Appendix \ref{sec:appendix_reflection_coefficent}. The reflection coefficient is plotted in Figure \ref{fig:reflection_coefficent} as a function of frequency.
For $\xi\gg1$, the equation for the reflection coefficient reduces to the following form,
\begin{equation}
    \label{eq:approx_RE}
    \lim_{\xi\rightarrow\infty}R_E=\left|\frac{a-1}{a+1}\right|^2.
\end{equation} 
Expanding equation \eqref{eq:r_e} about $\xi=0$ gives, to leading order,
\begin{equation}
    R_E=1-4\pi\frac{h\omega}{v_A}+O(\xi^2).
\end{equation}

The leakage timescale, $\tau_{leakage}$, which is the timescale at which wave energy in the corona is lost through leakage into the chromosphere, is given by
\begin{equation}
    \label{eq:tau_leakage}
    \tau_{leakage} = \frac{L}{v_A|\log R|}=\frac{2L}{v_A|\log R_E|}.
\end{equation}
This equation can be derived by considering a partially confined wave; its amplitude, $u_0$, behaves as
\begin{equation}
    \begin{aligned}
    u_0\propto R^{\lfloor v_At/L\rfloor}&=\exp\left(\left\lfloor \frac{v_At}{L}\right\rfloor\log R\right) \\
    &\approx\exp\left(-\frac{v_A|\log R|}{L}t\right),
    \end{aligned}
\end{equation}
where $\lfloor \rfloor$ denotes the floor function, which takes the integer part of the input. In other words, a factor $R$ of the wave amplitude will be lost each time it partially reflects off either end of the loop. In Section \ref{sec:damping_rate} we showed that the damping rate of a resonant field line can be approximated by equation \eqref{eq:dr_approx}, and therefore the resistive timescale is given by
\begin{equation}
    \label{eq:tau_resistive}
    \tau_{resistive}=\left[\frac{4}{3}\frac{(\eta+\nu)\omega^2}{v_A^2}(\nabla_\perp v_A)^2\right]^{-1/3}.
\end{equation}
Both timescales are plotted in Figure \ref{fig:leak_vs_resis_timescale}, where the following parameters were used: $\eta+\nu=1\si{.m^2.s^{-1}}$,  $v_A=1\si{.Mm.s^{-1}}$, $\nabla_\perp v_A=1\si{.s^{-1}}$, $L=100\si{.Mm,}$ and $h=150\si{.km}$, with $R_E$ being given by equation \eqref{eq:r_e}. It can be seen that for these parameters the leakage timescale is shorter than the resistive timescale. Therefore, leakage could play an important role in the dynamics. However, as shown in the introduction, we estimate that a viable heating mechanism needs to have a damping rate of about $10\si{.s^{-1}}$ which gives a timescale that is much shorter than the leakage timescale

Leakage acts to reduce the wave energy, but it also acts to reduce the heating rate and so it is perhaps not clear how leakage will affect the damping rate. The damping rate is plotted as a function of the reflection coefficient in Figure \ref{fig:dr_vs_leak} for a resonant field line. This was produced using equation \eqref{eq:dr} with the following parameters: $\eta+\nu=1\si{.m^2.s^{-1}}$,  $v_A=1\si{.Mm.s^{-1}}$, $\nabla_\perp v_A=1\si{.s^{-1}}$, and $f=10^{-2}\si{.Hz}$. It can be seen that increasing the leakage, $-\log R$, causes the damping rate, $\gamma$, to decrease. Therefore, leakage has the effect of reducing the damping rate. It can be seen that for $-\log R \ll L l_{ph}^{-1}$  the damping rate is largely independent of the reflection coefficient and can be approximated with equation \eqref{eq:dr_approx}. For $-\log R \gg L l_{ph}^{-1}$ the damping rate can be approximated by the blue dashed curve, which has the following equation,
\begin{equation}
    \label{eq:dr_high_leakage}
    \gamma \approx 2\frac{(\eta+\nu)\omega^2L^2}{v_A^4}(\nabla_\perp v_A)^2 \left(\frac{1+R}{1-R}\right)^2,
\end{equation}
which is derived in Appendix \ref{sec:dr_weak_reflection}. The horizontal red dotted line in Figure \ref{fig:dr_vs_leak}  was produced by replacing the cubic exponentials with linear exponentials (as in Section \ref{sec:damping_rate}); it can be seen to be independent of leakage. This suggests that it is because of the cubic nature of phase mixing that leakage causes the damping rate to decrease. More intuitively, the leakage prevents the waves phase-mixing down to very short length scales before they leak out of the loop. We have marked in magenta the value of the reflection coefficient for $f=10^{-2}\si{.Hz}$ and $h=150\si{.km}$. This shows that for this parameter space, the damping rate can be approximated by equation \eqref{eq:dr_high_leakage}.

\section{Multiple harmonics}
\label{sec:multiple_harmonics}

\begin{figure}
    \centering
    \includegraphics[width=0.4\textwidth]{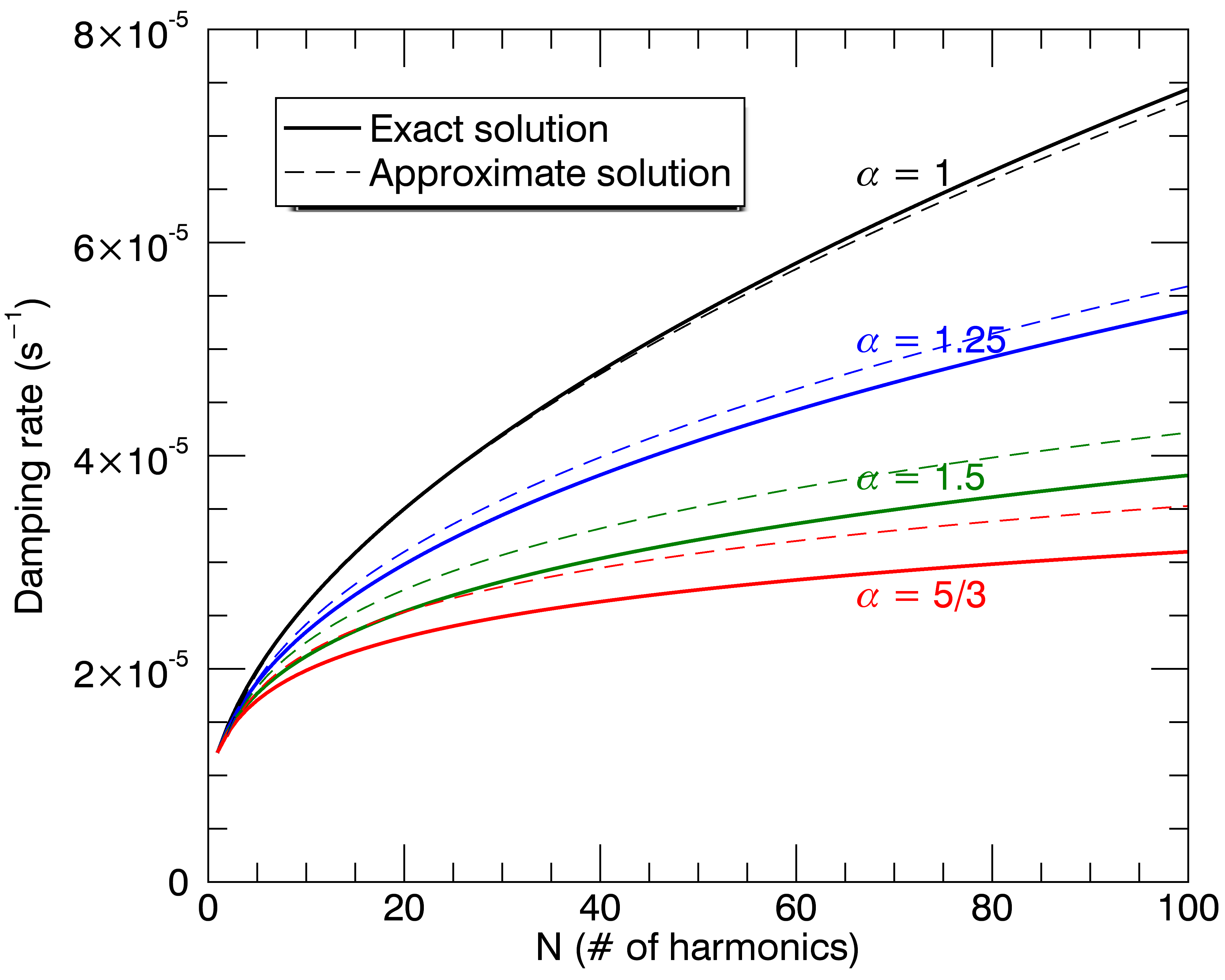}
    \caption{Plots of the damping rate, $\gamma$, of a resonant field line as a function of the number of excited harmonics. The approximate curves are produced by replacing the summation in equation \eqref{eq:dr_multi_harmonics} with an integral.}
    \label{fig:dr_vs_harmonic}
\end{figure}

As shown in the literature, for example by \citet{Morton2016} and \citet{Morton2019}, coronal loops do not oscillate at one frequency. Instead, they oscillate at a spectrum of frequencies. These latter authors show that velocity fluctuations in the corona approximately follow a power law with a bump at approximately five-minute periods which roughly correspond to the p-mode frequencies. For simplicity, we assume that our velocity fluctuations obey a single power law. Additionally, we drive with a series of sinusoidal drivers at the natural frequencies because a broadband driver drives resonances in a manner that is similar to a superposition of monochromatic drivers \citep{Wright1995}.

In Section \ref{sec:analytic_soln}, we show that for a driver of the form
\[u=u_0\sin(\omega t),\]
the steady-state solution for $u$ is given by equation \eqref{eq:u_steady_state}. Since we are assuming linear waves, we can easily find the full solution for a driver consisting of many harmonics,
\begin{equation}
    u=u_0\sum_{n=1}^N\frac{(n\omega_1)^{-\alpha/2}\sin(n\omega_1 t + \phi_n)}{\sqrt{\int_{-l}^lA_n^2+B_n^2dz}},
\end{equation}
which gives,
\begin{equation}
    \begin{aligned}
    u=u_0\sum_{n=1}^N(n\omega_1)^{-\alpha/2}&\left[\frac{A_n\sin(n\omega_1t+\phi_n)}{\sqrt{\int_{-l}^lA_n^2+B_n^2dz}}+\right. \\
    &\ \left.\frac{B_n\cos(n\omega_1t+\phi_n)}{\sqrt{\int_{-l}^lA_n^2+B_n^2dz}}\right],
    \end{aligned}
\end{equation}
\begin{equation}
    \begin{aligned}
    b=-u_0\sqrt{\mu\rho}\sum_{n=1}^N(n\omega_1)^{-\alpha/2}&\left[\frac{C_n\sin(n\omega_1t+\phi_n)}{\sqrt{\int_{-l}^lA_n^2+B_n^2dz}}+\right. \\
    &\ \left.\frac{D_n\cos(n\omega_1t+\phi_n)}{\sqrt{\int_{-l}^lA_n^2+B_n^2dz}}\right],
    \end{aligned}
\end{equation}
where $\omega_1$ is the fundamental angular frequency, $\phi_n$ is a random phase, $\alpha$ is a constant which controls the power spectrum of the driver, 
\begin{equation}
    A_n=\sum_{k=0}^\infty(-1)^kR^k\exp[-(z_k/l_{ph})^3]\cos(n\omega_1z_k/v_A),
\end{equation}
\begin{equation}
    B_n=-\sum_{k=0}^\infty(-1)^kR^k\exp[-(z_k/l_{ph})^3]\sin(n\omega_1z_k/v_A),
\end{equation}
\begin{equation}
    C_n=\sum_{k=0}^\infty R^k\exp[-(z_k/l_{ph})^3]\cos(n\omega_1z_k/v_A),
\end{equation}
\begin{equation}
    D_n=-\sum_{k=0}^\infty R^k\exp[-(z_k/l_{ph})^3]\sin(n\omega_1z_k/v_A).
\end{equation}
This solution has the property that
\begin{equation}
    \int_{-l}^l\rho\langle u^2\rangle dz=\frac{1}{2}\rho u_0^2\sum_{n=1}^N(n\omega_1)^{-\alpha},
\end{equation}
which can be shown by the fact that,
\begin{equation}
    \begin{aligned}
        \langle\sin&(n\omega t + \phi_n)\sin(m\omega t + \phi_m)\rangle \\
        &=\frac{n\omega}{2\pi}\int_0^{2\pi/(n\omega)}\sin(n\omega t + \phi_n)\sin(m\omega t + \phi_m)dt \\
        &=\begin{cases}
        1/2, & n=m, \\
        0, & \text{otherwise},
        \end{cases}
    \end{aligned}  
\end{equation}
for $n\le m\in\mathds{N}$. This means we can easily control the power spectrum of the steady-state kinetic energy of the system with $\alpha$. The average Poynting flux, $-B_0\langle ub\rangle/\mu$, is given by
\begin{equation}
    -B_0\langle ub\rangle/\mu=-\frac{1}{2}\rho u_0^2 v_A\sum_{n=1}^N(n\omega_1)^{-\alpha}\frac{A_nC_n+B_nD_n}{\int_{-l}^lA_n^2+B_n^2dz}.
\end{equation}
The damping rate, $\gamma$, is given by
\begin{equation}
        \label{eq:dr_multi_harmonics}
        \gamma =\frac{\sum_{n=1}^{N}n^{-\alpha}\gamma_n}{\sum_{n=1}^Nn^{-\alpha}},
\end{equation}
where $\gamma_n$ is given by
\begin{equation}
    \gamma_n=v_A\frac{[A_nC_n+B_nD_n]_{-l}^l}{\int_{-l}^lA_n^2+B_n^2dz}.
\end{equation}
Equation \eqref{eq:dr_multi_harmonics} takes the form of a weighted average of the set $\{\gamma_n\}$ with weights $\{n^{-\alpha}\}$. The damping rate is plotted in Figure \ref{fig:dr_vs_harmonic} as a function of the number of excited harmonics for different values of $\alpha$ using the following parameter values: $\eta+\nu=1\si{.m^2.s^{-1}}$, $v_A=1\si{.Mm.s^{-1}}$, $\nabla_\perp v_A=1\si{.s^{-1}}$, and $L=100\si{.Mm}$, corresponding to a fundamental harmonic of $f_1=5\times10^{-3}\si{.Hz}$. As expected, increasing the number of harmonics increases the damping rate. Using equation \eqref{eq:dr_approx} as an approximation for $\gamma_n$ and replacing the summation with an integral gives
\begin{equation}
    \begin{aligned}
        \gamma &\approx \left(\frac{4}{3}\frac{(\eta+\nu)\omega_1^2}{v_A^2}(\nabla_\perp v_A)^2\right)^{1/3}\frac{\int_{n=1}^Nn^{-(\alpha-2/3)}dn}{\int_{n=1}^Nn^{-\alpha} dn} \\
        &=\gamma_1^{1/3}\frac{\alpha-1}{5/3-\alpha}\frac{N^{5/3-\alpha}-1}{1-N^{-(\alpha-1)}},
    \end{aligned}
\end{equation}
for $\alpha\ne5/3$ and $\alpha\ne1$.
For $\alpha=5/3,$ it is approximated by
\begin{equation}
    \gamma\approx\gamma_1\frac{2}{3}\frac{\log N}{1-N^{-2/3}},
\end{equation}
and for $\alpha=1$ it is approximated by
\begin{equation}
    \gamma\approx\gamma_1\frac{3}{2}\frac{N^{2/3}-1}{\log N}.
\end{equation}
These approximations are also plotted as dashed curves in Figure \ref{fig:dr_vs_harmonic}. The reason there is a noticeable error between the exact and approximate solutions in Figure \ref{fig:dr_vs_harmonic} is twofold: first, the approximate solution approximates $\gamma_n$ using equation \eqref{eq:dr_approx}, and second, the summations in equation \eqref{eq:dr_multi_harmonics} have been replaced with an integral. It can be seen that for larger $\alpha,$ increasing $N$ does relatively little to change the damping rate. Figure \ref{fig:dr_vs_harmonic} uses a range of $\alpha$ values including 1.5 and $5/3$ as these are the values predicted from MHD turbulence theory \citep{Bruno2016}. \citet{Morton2016} provides observations of the power spectra of velocity fluctuations in the quiet sun, the active regions, and the coronal holes. 
These latter authors find that the slope varies from $\alpha=1$ to $\alpha=1.53$ for higher frequencies, although they are only able to measure up to frequencies of around $10^{-2}\si{.Hz}$. \citet{Podesta2007} measure the power spectra in the solar wind and can measure up to $10^{-1}\si{Hz}$ and find the slope to be between $\alpha=1.5$ and $\alpha=5/3$. From Figure \ref{fig:dr_visc_vs_phase}, it can be seen that for the higher frequencies, the heating due to parallel gradients start to dominate. Therefore, if we use a value of $N$ of higher than 100, we can see that parallel gradients will start to dominate the heating. Therefore, we can no longer describe the system as being heated mainly by phase mixing.

\section{Conclusions}

We believe an upper bound for the damping rate of laminar phase-mixed Alfv\'en waves oscillating with a given power spectrum is given by equation \eqref{eq:dr_multi_harmonics} with $\gamma_n$ approximated by \eqref{eq:dr_approx}, provided our assumptions (see Section \ref{sec:justification}) are valid. Here, we provide a brief argument for why we believe this to be an upper bound: In Section \ref{sec:damping_rate}, we show that the damping rate increases with time but converges towards a maximum as the system approaches steady state (the blue curve illustrates this in Figure \ref{fig:en_int_power_over_en_wave}) and our equations were derived by assuming the system had reached a steady state. This is important because (as pointed out by \citet{Arregui2015}) waves may not have a chance to reach steady state. Also, we show that the damping is largest at the natural frequencies (illustrated in Figure \ref{fig:dr_vs_freq}) and our equations were derived by only exciting the resonant frequencies with the driver. In Section \ref{sec:leakage}, we show that allowing waves to leak through the atmosphere decreases the damping rate and equation \eqref{eq:dr_approx} was derived assuming perfect reflection at the transition region. We have not considered the thermodynamics, but \citet{Cargill2016} showed that the thermodynamic response due to heating reduces the density gradients, which reduces the damping rate. As stated in Section \ref{sec:equations}, we solve the linearised MHD equations. \citet{Prokopyszyn2019} found that provided there was no turbulence (which is ensured by the presence of an ignorable coordinate and the fact that MHD turbulence of pure Alfv\'en waves is a strictly 3D phenomenon (see \citet{Howes2013})) then for $u/v_A\lessapprox0.1$ the non-linearities have a negligible effect on the damping rate. One problem with the results from \citet{Prokopyszyn2019} is that an unphysically large value for the dissipative coefficients had to be used due to numerical constraints.  Non-linearities can trigger turbulence, for example through the interaction of counter-propagating Alfv\'en waves (\citet{Hollweg1986}, \citet{vanBallegooijen2011} and \citet{Shoda2019}) or via the Kelvin-Helmholtz / tearing mode instability (\citet{Heyvaerts1983} and \citet{Browning1984}). Turbulence leads to the transfer of energy into higher wavenumbers/frequencies, which causes the damping rate to increase. However, our claim is that equation \eqref{eq:dr_multi_harmonics} is an upper bound for laminar waves.
Finally, \citet{Threlfall2011} found that Hall MHD terms produce wave dispersion and reduce the damping rate.

Provided equation \eqref{eq:dr_multi_harmonics} is a valid upper bound for the damping rate, this implies that phase mixed Alfv\'en waves at observed frequencies are unlikely to play a role in coronal heating. This can be seen from Figure \ref{fig:dr_vs_harmonic} which predicts a damping rate of no more than $10^{-4}\si{.s^{-1}}$.  We believe a damping rate of the order $10^{-1}\si{.s^{-1}}$ is required to heat the corona  (as shown in the introduction). Changing our parameters will change the damping rate. The damping rate may be high enough in some locations of the coronal atmosphere, for example near null points where the cross-field viscosity becomes much stronger. In this study it is assumed that the flow remains laminar. \citet{Browning1984} show that a phase-mixed standing Alfv\'en wave will trigger the Kelvin-Helmholtz instability at its antinode, which could lead to a turbulent cascade, and these latter authors calculated that this could lead to sufficient heating in the corona. They find the instability is triggered at the antinodes of the wave as this is where the magnetic field is smallest. In theory, multiple harmonics should help to stabilise the field as the magnetic field perturbations should be more uniformly distributed across the field line if this is the case. Future work could investigate the effects of multiple harmonics on the phase-mixing-induced Kelvin-Helmholtz instability.

\begin{acknowledgements}
We thank Andrew Wright and Ineke De Moortel for help with checking through the paper and useful discussions. This project has received funding from the Science and Technology Facilities Council (U.K.) through the consolidated grant ST/N000609/1.
\end{acknowledgements}

\begin{appendix}

\section{Reflection  coefficient}
\label{sec:appendix_reflection_coefficent}

\citet{Hollweg1984resonances} derived an expression for the reflection coefficient (equation \eqref{eq:r_e_hollweg}). He derived this by splitting his domain into two regions, $z<0$ corresponds to the chromosphere with an exponentially growing Alfv\'en speed with pressure scale height, $h$, and $z>0$ corresponds to the corona with a uniform Alfv\'en speed. By considering a source of waves in $z>0$ he calculates that in general, the waves have the form
\begin{equation}
    u = 
    \begin{cases}
        C_1H_0^{(2)}(\xi)e^{i\omega t}, & z<0, \\
        C_2e^{i\omega (t - s/v_A)}+C_3e^{i\omega (t + s/v_A)}, & z>0,
    \end{cases}
\end{equation}
\begin{equation}
    b=
    \begin{cases}
    -i\frac{C_1}{v_A}B_0H_1^{(2)}(\xi)e^{i\omega t},& z<0, \\
    -\frac{B_0}{v_A}[C_2e^{i\omega (t - s/v_A)}-C_3e^{i\omega (t + s/v_A)}],& z>0,
    \end{cases}
\end{equation}
where $C_1$, $C_2$, $C_3$ are arbitrary constants and $\xi$ is given by equation \eqref{eq:xi_hollweg}. We modify these equations slightly because we assume there is a discontinuous jump in Alfv\'en speed of a factor $a$ from the chromosphere to the the corona due to the jump in density. Taking $v_A$ as the Alfv\'en speed in the corona and $v_A/a$ as the speed at the top of the chromosphere then this gives the same expression for $u$ but with $b,$ now given by
\begin{equation}
    b=
    \begin{cases}
    -i\frac{C_1a}{v_A}B_0H_1^{(2)}(\xi)e^{i\omega t},& z<0, \\
    -\frac{B_0}{v_A}[C_2e^{i\omega (t - s/v_A)}-C_3e^{i\omega (t + s/v_A)}],& z>0,
    \end{cases}
\end{equation}
and $\xi$ given by \eqref{eq:xi}. Despite $v_A$ being discontinuous, it can be shown from Farday's, Amp\`ere's, and Ohm's law that $u$ and $b$ must be continuous \citep{Hollweg1984simple}. The continuity conditions can be used to eliminate $C_1$ and the reflection coefficient, $R$, is given by $C_2/C_3$. Hence,
\begin{equation}
    \tag{\ref{eq:r_e}}
    R_E = \left|\frac{H_0^{(2)}(\xi)+iaH_1^{(2)}(\xi)}{H_0^{(2)}(\xi)-iaH_1^{(2)}(\xi)}\right|^2.
\end{equation}

\section{$\gamma$ in the high leakage limit}

\label{sec:dr_weak_reflection}
In the high leakage limit, equation \eqref{eq:full_soln} reduces to 
\begin{equation}
    u=u_0\sum_{k=0}^\infty(-1)^kR^kH(\theta_k)\sin(\omega\theta_k).
\end{equation}
Letting $t\rightarrow\infty$, considering only the apex of the loop ($z=0$), and replacing the sine function with a complex exponential gives
\begin{equation}
    \label{eq:full_soln_large_leakage0}
    u=u_0\exp[i\omega(t-l/v_A)]\sum_{k=0}^\infty(-1)^k[R\exp(-2i\omega l / v_A)]^k.
\end{equation}
This equation takes the form of a geometric series which can be evaluated, provided $R<1$, to give
\begin{equation}
    \label{eq:full_soln_large_leakage}
    u=u_0\frac{\exp[i\omega(t-l/v_A)]}{1+R\exp(-i\omega L / v_A)},
\end{equation}
where $L=2l$.
The imaginary part of equation \eqref{eq:full_soln_large_leakage} has maxima if odd harmonics are excited, i.e. $\omega=(2n+1)\pi v_A/L$, $n\in\mathds{N}$ then  $\exp(-i\omega L/v_A)=-1$. The gradient of equation \eqref{eq:full_soln_large_leakage} perpendicular to the field for $\omega=(2n+1)\pi v_A/L$ is given by
\begin{equation}
    \frac{\partial u}{\partial x}=i\frac{\omega l}{v_A^2}\frac{d v_A}{d x}\frac{1+R}{(1-R)^2}u_0\exp[i\omega(t-l/v_A)].
\end{equation}
Therefore, the damping rate can be approximated with $\omega=(2n+1)\pi v_A / L$ to give
\begin{equation}
    \tag{\ref{eq:dr_high_leakage}}
    \begin{aligned}
        \gamma&\approx(\eta+\nu)\frac{\langle |du/dx|^2\rangle}{\langle |u|^2\rangle} \\
        &=2(\eta+\nu)\frac{\omega^2L^2}{v_A^4}(\nabla_\perp v_A)^2\left(\frac{1+R}{1-R}\right)^2.
    \end{aligned}
\end{equation}

\end{appendix}

\bibpunct{(}{)}{;}{a}{}{,} 
\bibliographystyle{aa}        
\bibliography{bibliography.bib}           

\end{document}